\documentclass[prl,twocolumn,showpacs,superscriptaddress]{revtex4}
\usepackage{graphicx}
\usepackage{dcolumn}
\begin{document}
\title{Identifying and Indexing Icosahedral Quasicrystals from Powder
Diffraction Patterns}
\author{Peter J. Lu}
\altaffiliation{Current address: Department of Physics, Harvard
University, Jefferson Laboratory, 17 Oxford Street, Cambridge, MA
02138.} \email{plu@fas.harvard.edu} \affiliation{Department of
Physics, Princeton University, Princeton, NJ 08544}

\author{Kenneth Deffeyes}
\affiliation{Department of Geology, Princeton University,
Princeton, NJ 08544}

\author{Paul J. Steinhardt}
\affiliation{Department of Physics, Princeton University,
Princeton, NJ 08544}

\author{Nan Yao}
\affiliation{Princeton Materials Institute, Princeton University,
Princeton, NJ 08544}

\date{August 15, 2001}

\begin{abstract}
We present a scheme to identify quasicrystals based on powder
diffraction data and to provide a standardized indexing. We apply
our scheme to a large catalog of powder diffraction patterns,
including natural minerals, to look for new quasicrystals. Based
on our tests, we have found promising candidates worthy of further
exploration.
\end{abstract}

\pacs{61.10.Nz, 61.44 Br}
\maketitle

Quasicrystals are solids whose lattices exhibit a rotational
symmetry, such as five-fold symmetry, that is forbidden for
periodic crystals.\cite{steinhardt1} Because the atoms are
arranged quasiperiodically, the structure's symmetry is not
subject to the usual mathematical restrictions of crystallography.
Quasiperiodic translational order has physical consequences. For
example, since electrons and phonons in quasicrystals do not
encounter a periodic potential, quasicrystals have unusual
resistive and elastic properties, and these have been exploited in
several applications.\cite{goldman1} To date, known quasicrystals
have been found by serendipity or by probing stoichiometric
variations around other already-known quasicrystals. Furthermore,
all known quasicrystals are synthetic; no natural quasicrystal has
ever been identified. A more systematic way to search for
quasicrystals, including natural quasicrystals, is desirable, and
one way is to search through collections of diffraction data.
Although a two-dimensional electron diffraction pattern would
immediately show a quasicrystal's salient forbidden symmetry, no
large collection of such patterns exists. However, a collection of
over eighty thousand \textit{powder} diffraction patterns in
digital form, the Powder Diffraction File (ICDD-PDF), is published
by the International Center for Diffraction Data. The catalog
contains synthetic inorganic and organic phases, as well as
nine-thousand mineral patterns. Because the powder diffraction
pattern of a material averages over all orientations, only the
magnitude (and not the direction) of the scattering vector in
reciprocal space is preserved, and a quasicrystal's distinctive
non-crystallographic symmetry cannot be observed directly.
\textit{A priori}, it is unclear if quasicrystals can be
identified from their powder diffraction patterns alone.

In this paper, we present a method to identify, classify and index
icosahedral quasicrystals based solely on their powder diffraction
patterns. We apply the method to the ICDD-PDF and find the
best-fit quasicrystal candidates. Our automated procedure picks
out the known quasicrystals in the ICDD-PDF and, for the cases in
which indexing has been published, produces the same indices.
Although the remainder of the catalog is supposed to consist of
periodic crystals, it was assembled over many years, including
decades prior to the 1984 discovery of quasicrystals.
Consequently, it is conceivable that the catalog includes some
quasicrystals that were never identified as such. Based on our
studies, we report promising materials.

The diffraction pattern of an ideal three-dimensional quasicrystal
consists of Bragg peaks located on a lattice given by $\vec{Q} =
\sum_{i = 1}^6 n_i \vec{b}_i \label{eqn:Qvector}$, where the
$\vec{b}_i$ are basis vectors pointing to the vertices along the
six five-fold symmetry axes of a regular icosahedron in three
dimensions, and the $n_i$ are integers that index each vector. The
quasicrystal formula is similar to that for a crystal except that
the number of basis vectors is greater than the number of
dimensions (three), a consequence of the non-crystallographic
symmetry. We choose the $\vec{b_i}$ following the convention in
\cite{janot1}:
\begin{eqnarray}
\vec{b}_1 = (1, \tau, 0),\ \vec{b}_2 & = & (\tau, 0, 1),\
\vec{b}_3 = (0, 1, \tau), \\ \vec{b}_4 = (-1, \tau, 0),\ \vec{b}_5
& = & (\tau, 0, -1),\ \vec{b}_6 = (0, -1, \tau) \nonumber
\end{eqnarray}
where $\tau$ is the golden ratio, $(1+\sqrt{5})/2$. An equivalent
way to index the position of the three-dimensional
reciprocal-space scattering vector is to use a scheme analogous to
crystallographic Miller $(h\ k\ l)$ indices. Six integer indices
grouped into pairs describe the distance along Cartesian basis
vectors, $\vec{Q} = (h+h' \tau)\hat{x} + (k+k'\tau)\hat{y} + (l+l
'\tau)\hat{z}$. The $(h/h'\ k/k'\ l/l')$ indices are permutations
of the $n_i$:
\begin{eqnarray}
h = n_1 - n_4 , \ & \ h' = n_2 + n_5 , \ & \ k = n_3 - n_6, \\ k'
= n_1 + n_4 , \ & \ l = n_2 - n_5 , \ & \ l' = n_3 + n_6 \nonumber
\end{eqnarray}
Because the $(h/h'\ k/k'\ l/l')$ indices express distances along
Cartesian axes, the advantages of orthogonal coordinate axes are
conveniently recovered. Also, associated with every $\vec{Q}$ is a
vector $\vec{Q}_\perp$, constructed from another integer linear
combination of the same basis vectors: $\vec{Q}_\perp =
(h'-h\tau)\hat{x} + (k'-k\tau)\hat{y} + (l'-l\tau)\hat{z} =
\sum_{i = 1}^6 n_i \vec{b}^\perp_i$, where
\begin{eqnarray}
\vec{b}^\perp_1 = (-\tau, 1, 0),\ \vec{b}^\perp_2 & = & (1, 0,
-\tau),\ \vec{b}^\perp_3 = (0, -\tau, 1), \\ \vec{b}^\perp_4 =
(\tau, 1, 0),\ \vec{b}^\perp_5 & = & (1, 0, \tau),\
\vec{b}^\perp_6 = (0, \tau, 1) \label{eqn:perpbasisvectors} \nonumber
\end{eqnarray}
In powder diffraction patterns, the reciprocal space is collapsed
to one dimension, where all vectors with the same magnitude
$|\vec{Q}|$ are degenerate. The magnitude $|\vec{Q}|\equiv Q$ is
the reciprocal of the d-spacing, the quantity listed in the
entries of the ICDD-PDF. $Q^2$ and $|\vec{Q}_{\perp}|^2 \equiv
Q^2_\perp$ can now be expressed as integer linear combinations of
only two basis vectors, whose lengths are related by the golden
ratio: $Q^2 \propto N + \tau M$ and $Q_\perp^2 \propto N \tau -
M$, where $N = 2 \sum_{i=1}^6 n^2_i$ and $M = h'^2 + k'^2 + l'^2 +
2(hh' + kk' + ll')$. Each peak in a quasicrystal powder
diffraction pattern can be indexed by the two integers $N$ and
$M$. The entire diffraction pattern can be scaled by a factor of
$\tau^3$ along the $Q$ direction, and the support will remain the
same, a consequence of the self-similarity of the pattern. That
is, for each lattice vector at position $Q^2 = (N,M)$, there is
another lattice vector of similar intensity at position
$(N',M')=\tau^6 Q^2$. Using the relation $\tau^2 = \tau + 1$,
\begin{eqnarray}
&(N',M') \equiv N' + M' \tau = \tau^6(N + M \tau) = &
\label{eqn:tauscaling} \\& (5M + 8N) + (8M + 13N)\tau \equiv (5M +
8N, 8M + 13N) & \nonumber
\end{eqnarray}
Although three other different indexing schemes persist in the
literature to describe icosahedral quasicrystal powder
patterns,\cite{bancel1, cahn1, elser1} they are all analytically
equivalent to the convention given here.

There are three distinct icosahedral reciprocal lattices: simple
icosahedral (SI), face-centered icosahedral (FCI) and
body-centered icosahedral (BCI). The SI, FCI and BCI reciprocal
lattices correspond to real-space lattices obtained by, for
example, placing identical ``atoms'' at each lattice point of a
simple-hypercubic, face-centered hypercubic and body-centered
hypercubic lattice in six dimensions, respectively, and projecting
down to three dimensions. (In \cite{mermin1}, the SI, FCI and BCI
lattices are referred to as P*, I* and F*, respectively.) For
these primitive lattices, the intensity is given
by:\cite{steinhardt2}
\begin{equation}
I \propto \left[ {\sin(Q_\perp / 2) \over Q_\perp / 2} \right]^2
\label{eqn:intensity}
\end{equation}
Note that $I$ increases as $Q_\perp \rightarrow 0$ so that bright
peaks---the ones likely to be observed experimentally---have small
$Q_\perp$. Eq.~(\ref{eqn:intensity}) does not account for chemical
and other effects that modulate the intensities in the diffraction
pattern of a real material, but it provides a guide as to which
peaks should be observed when testing real patterns.

\textit{Testing patterns.} The first step in testing a real powder
pattern is to find the best possible peak-by-peak match between
that pattern $\{Q_i\}$ and the perfect quasicrystal template
pattern $\{q_i\}$. Then, various statistical tests can be applied
to compare that match to what is expected for a true quasicrystal.

If we were to match to a periodic crystal template pattern, the
test would be more straightforward. For any finite interval of $Q$
over which a pattern is measured, a real crystal has a finite
number of diffraction peaks. One also has good \textit{a priori}
estimates for the nonzero minimum distance between nearest peaks,
which is set by the lattice constant. These features, useful in
matching the template to the real pattern, allow a unique
indexing.

For the quasicrystal, the process is more complicated because the
number of peaks in any finite interval of the ideal pattern is
infinite (the pattern is dense) and, consequently, there is no
uniquely defined lattice constant or indexing. How does one
sensibly match a real pattern, with a finite number of peaks in a
given interval $\Delta Q$, to a quasicrystal template pattern,
with a dense set of peaks in that same interval? One must account
for the intensity of the peaks, not just their $Q$. Although the
quasicrystal powder pattern is dense, most of its peaks within any
finite interval of $Q$ have a large $Q_\perp$ and are predicted by
Eq.~(\ref{eqn:intensity}) to be too dim to distinguish from the
noise present in any real experiment. So, instead of finding the
template $q_{i'}$ which comes closest to the observed $Q_i$ (i.e.
minimizing $|Q_i^2 - q_{i'}^2|$ alone), we instead minimize
$|Q_i^2 - q_{i'}^2 | / I_{i'}$, which includes the intensity
$I_{i'}$ of the $i'$th template peak. This approach tends to match
real $Q_i$ to the nearest \textit{bright} peaks in the template
pattern, naturally accounting for phason shifts, imperfections and
experimental error that may shift $Q_i$ from its ideal value and
cause an incorrect assignment to some $q_{i'}$ with an
unrealistically small intensity.

The procedure for finding the best match between the real and
template patterns, then, is as follows. Choose some bright peak in
the ideal pattern $(N_0, M_0)$ with a low value of $N$. An
``attempted match'' consists of re-scaling its magnitude to match
the first real peak in the pattern, $Q_1$. Identify each remaining
peak $Q_{i \neq 1}$ in the real pattern with the template peak
$q_{i'}$ which minimizes $|Q_i^2 - q_{i'}^2 | / I_{i'}$. Next,
introduce a goodness-of-fitness parameter $S_1$ to characterize
the attempted match between the $\{Q_i\}$ and the $\{q_{i'}\}$.
Repeat the process by assigning $(N_0,M_0)$ to the second real
peak $Q_2$ and compute its goodness-of-fit, $S_2$. After repeating
for each real peak, use the lowest $S_j$ to decide the best
overall match between the real pattern and template. Each real
peak $Q_i$ is now assigned the set of indices $\{n_{i'}\}$ or,
equivalently, $\{ N_{i'}, M_{i'} \}$ of the matching template
peak.

The goodness-of-fit parameter $S_j$ depends on $\bar{\Delta} =
\langle | Q_i - q_{i'} | / Q_i \rangle$ where $\langle {\cal O}_i
\rangle$ denotes an intensity-weighted average $(\sum_i \sqrt{I_i}
{\cal O}_i) /(\sum_i \sqrt{I_i})$. $\bar{\Delta}$ measures the
fractional deviation between $Q_i$ and $q_{i'}$ and is weighted by
intensity for the same reasons as above. (Results do not change
significantly if $\sqrt{I_i}$ is replaced by $I_i$ or some similar
function of intensity.) One challenge is that for each good fit of
$Q_i$ to a template peak $q_{i'}$ labelled by $(N,M)$, there is
also an equally good fit to the peak $\tau^{6k}(N,M)$ where $k$ is
any integer, due to the self-similarity of the lattice; see
Eq.~(\ref{eqn:tauscaling}). This leads to the practical problem of
deciding which match to choose and also the annoyance that, unlike
a crystal, the indexing is not unique. To produce a unique and
sensibly standardized indexing, we simultaneously minimize
$\bar{N} = \langle N_i \rangle$. The goodness-of-fit $S$ can be
taken to be a linear combination of these two parameters, $S = a
\bar{\Delta} + \bar{N}$. The results are relatively insensitive to
the choice of $a$ (up to an order of magnitude) provided that the
$\bar{\Delta}$ and $\bar{N}$ contributions to $S$ are both
non-negligible for known quasicrystals; typical values are
$\bar{N}\approx 30$ and $\bar{\Delta}\approx 0.3$, and we used
$a=500$ in all data runs.

\begin{figure}
\begin{center}
\includegraphics*{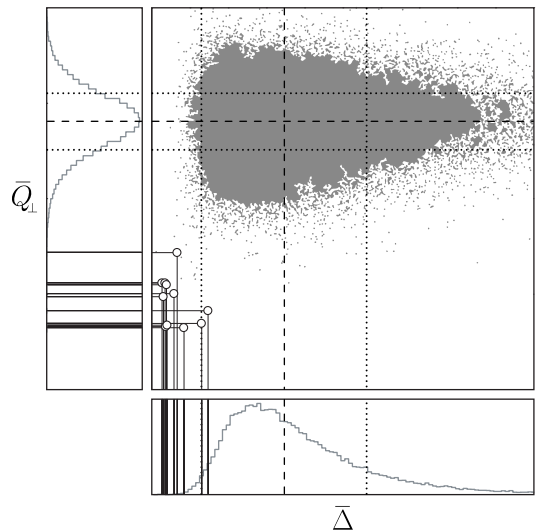}
\end{center}
\caption{A plot of the distribution of quantities $\bar{\Delta}$
and $\bar{Q_\perp}$ for eleven patterns identified as FCI
quasicrystals in the ICDD-PDF (white circles) and 60,000 patterns
identified as crystals (grey dots). The large square shows the
scatter plot in the two-dimensional parameter space, in arbitrary
units with the origin at lower left. Rectangles along either axis
show the individual histograms of $\bar{\Delta}$ and
$\bar{Q}_\perp$. Means and standard deviations for each individual
quantity are marked by dashed and dotted lines, respectively. The
quasicrystals are projected to show that they lie at the tail of
the distributions.} \label{fig:scatterplot}
\end{figure}

If the real material is a crystal, the best attempted match to the
quasicrystal template (corresponding to the lowest $S_j$) is still
a poor match relative to a true quasicrystal, so we next introduce
statistical tests to measure the quality of the match. These tests
were found empirically by applying them first to known
quasicrystals, and they involve calculating several quantities for
each pattern. The first two statistical quantities, $\bar{\Delta}$
and $\bar{Q}_{\perp}$ (the intensity-weighted average of
$Q_\perp$), are discussed above. Quasicrystals have low values of
both, representing closer matches to brighter peaks. However,
finite resolution limits the number of peaks present above the
noise level; while these two parameters clearly separate out FCI
quasicrystals (see Fig.~\ref{fig:scatterplot}), they fail for the
SI case, where 1700 patterns score better than known SI
quasicrystals when ranking only by these two statistics. We
therefore consider another quantity that involves interrelations
among the bright peaks in a quasicrystal powder pattern.

Each peak in a powder pattern has a parity determined by $\eta
\equiv \sum_i n_i$. Even (odd) $\eta$ corresponds to  even (odd)
parity. Over a finite interval, certain sequences of even and odd
peaks appear. For a given bright peak at $Q_0$, there are other
bright peaks at $Q = Q_0 + \Delta Q$, where $\Delta Q \equiv
\Delta N + \Delta M \tau \equiv (\Delta N, \Delta M)$. All these
peaks should have low values of $Q_\perp$, and will therefore be
separated by small $\Delta Q_\perp^2 = \Delta N \tau - \Delta M$.
$\Delta M / \Delta N$ should approximate $\tau$, the golden ratio,
and as is well known, the best approximant is a ratio of
subsequent Fibonacci numbers. Hence, for every peak in the powder
pattern, we look for other peaks separated by ``Fibonacci
intervals," $(\Delta N, \Delta M)$, where $\Delta N$ and $\Delta
M$ are either successive Fibonacci integers or constant multiples
of these integers. In the SI case, we search for two types of
intervals: those occurring between peaks of opposite parity, which
involve successive Fibonacci numbers, such as $(2,3)$ and
$(34,55)$; and those occurring between peaks of the same parity,
which involve four times successive Fibonacci numbers, such as
$(8,12)$ and $(20,32)$. In the FCI and BCI cases, only even-parity
peaks are present, and we seek only sequences with four times
successive Fibonacci integers.

For each $i$th peak in the real pattern, we count the number of
other real peaks  that are separated from it by one of the
Fibonacci intervals. If there is only one such peak, then the
$i$th peak is considered part of a pair (or $2$-plet); if two,
part of a $3$-plet; etc. We define $M^{(m)}_i$ to be $1$ if the
$i$th peak is part of an $m$-plet and zero otherwise. Note that
$M^{(m)}_i = 1$ implies $M^{(m-1)}_i=1$. For example, if the ninth
peak in a pattern is separated from three other peaks by Fibonacci
intervals, then $M^{(2)}_9 = M^{(3)}_9 = M^{(4)}_9 = 1$ and
$M^{(5)}_9=0$. The values of $M^{(m)}_i$ for each peak are
combined into intensity-weighted averages $\bar{M}^{(m)} = \langle
M^{(m)}_i \rangle$.

The three average quantities $\bar{\Delta}$, $\bar{Q}_\perp$ and
$\bar{M}^{(4)}$ are unified into a single $\chi^2$ statistic,
which quantifies the degree of quasicrystallinity of a powder
X-ray diffraction pattern. First, the distribution for each
quantity, calculated for all sixty thousand entries in the
ICDD-PDF with a dozen peaks or more, is fit to a standard gaussian
by matching percentiles. Then, $\chi^2$ is calculated from the
three normalized measures. All patterns with any
worse-than-average quantity are discarded and the remaining
patterns with a high value of $\chi^2$ represent patterns with
characteristics most like those of the known quasicrystals.

\textit{Results.} The $\chi^2$ statistic has proven to be a
reliable method of identifying and indexing quasicrystal
diffraction patterns. For example, Fig.~\ref{fig:scatterplot}
shows a scatter plot of $\bar{\Delta}$~vs.~$\bar{Q}_{\perp}$ for
60,000 patterns from the ICDD-PDF. The eleven patterns known to be
synthetic quasicrystals of the FCI structure (by checking the
references given in the ICDD-PDF) are represented by white
circles; the remaining patterns are shown in grey. The clustering
of the eleven quasicrystals apart from the remaining points
demonstrates the success of the tests in separating out
quasicrystal patterns. The few grey dots around the white
quasicrystal circles are patterns identified as crystalline in the
ICDD-PDF that might be misidentified quasicrystals. However, many
of these possibilities are eliminated when the remaining
$\bar{M}^{(4)}$ test is applied. Because no one or two tests is
completely effective, we combine three quantities into a $\chi^2$
statistic.

Table~\ref{table:results} lists the top quasicrystal candidates
for the SI, FCI and BCI structures. For each structure, the list
includes a typical example of a known quasicrystal (except BCI,
where none is known), the top five synthetic materials identified
as crystalline, and the top mineral candidate in the ICDD-PDF. The
ranking should not be taken as an absolute measure. Different
versions of the statistical tests can alter the ranks of some
promising candidates, sometimes significantly, but those shown
remain among the top-ranked throughout. Our quasicrystal example
in the SI case is the recently discovered binary alloy,
CdYb.\cite{tsai1} Interestingly, a stoichiometrically similar
phase, (ostensibly crystalline) Cd$_6$Yb, placed seventh in the
BCI top candidate list. For the BCI case, there is no known
quasicrystal, and though the top candidates have high values of
$\chi^2$, the $\chi^2$ values cannot be compared directly between
different structures since statistical tests are different.
Nevertheless, the high scores suggest followup studies of the best
candidates, particularly InPtU and Co$_4$Sn$_{13}$Tb$_3$.

\begin{table}
\caption{Top quasicrystal candidates for SI, FCI, or BCI lattices.
Type refers to whether candidates were labeled in the ICDD-PDF as
quasicrystal, synthetic crystal or mineral crystal. Shown are a
typical quasicrystal, the top five synthetic and top mineral
candidates. The best known quasicrystals have $\chi^2$ values up
to $154.8$ (SI) and $397.8$ (FCI).}\label{table:results}
\begin{tabular}{c|c|c|c|c}
PDF Num & Formula & Lattice & Type & $\chi^2$ \\
\hline
New Data\cite{tsai1} & CdYb & SI & QC & 34.7 \\
27-901 & SnTe$_3$O$_8$ & SI & Syn & 28.4 \\
42-842 & InP$_3$ & SI & Syn & 27.9 \\
44-583 & CaUO$_4$ & SI & Syn & 27.4 \\
21-117 & Cd(MnO$_4$)$_2\cdot$6H$_2$O & SI & Syn & 21.4 \\
38-923 & K$_2$NaPdF$_6$ & SI & Syn & 19.8 \\
25-298 & Aktashite, Cu$_6$Hg$_3$As$_4$S$_{12}$ & SI & Min & 17.9 \\
\hline
48-1437 & Al$_{68.5}$Pd$_{22.1}$Mn$_{9.4}$ & FCI & QC & 92.0 \\
40-106 & Ba$_3$La$_{40}$V$_{12}$O$_{93}$ & FCI & Syn & 40.3 \\
19-261 & CaYb$_2$O$_4$ & FCI & Syn & 38.9 \\
31-1420 & UO$_3$ & FCI & Syn & 34.1 \\
41-979 & Cr$_{0.9}$Ta$_{5.1}$S & FCI & Syn & 33.8 \\
27-863 & Sr$_7$Y$_{13}$O$_4$(PO$_4$)$_{3}$(SiO$_4$)$_9$ & FCI & Syn & 33.7 \\
2-691 & Tantalite, (Fe,Mn)Ta$_2$O$_6$ & FCI & Min & 25.2 \\
\hline
21-379 & InPtU & BCI & Syn & 40.1 \\
50-1135 & Co$_4$Sn$_{13}$Tb$_3$ & BCI & Syn & 36.2 \\
42-1163 & Pb$_{10}$Al$_2$F$_{25}$Cl & BCI & Syn & 26.8 \\
45-1164 & Al$_{20}$Mo$_{1.656}$Th & BCI & Syn & 25.9 \\
36-1231 & LaNiO$_2$ & BCI & Syn & 24.9 \\
44-1412 & Gratonite, Pb$_9$As$_4$S$_{15}$ & BCI & Min & 14.3 \\
\end{tabular}
\end{table}

A further success is that our automated indexing of peaks for the
known quasicrystals matches the available published indices. In
some cases, while the published indices are aided by
two-dimensional diffraction data, our automated procedure uses
powder data only. Given the unique challenges in quasicrystal
indexing, described earlier, the successful indexing suggests that
our matching of template and real patterns automatically
incorporates standards obtained by individual analysis. Hence, as
more quasicrystals are discovered and added to catalogs, our
procedure could provide a standardized indexing.

Future work will proceed in several directions. A number of
materials merit further study, and, intriguingly, the most
promising examples are in the BCI class where no quasicrystal has
yet been found. Systematic studies of synthetic materials with
nearby stoichiometries may also be merited. Finally, we would like
to collect the powder diffraction patterns of other materials,
including known quasicrystals and crystal approximant phases, to
further refine our tests. We are interested in collaborating in
exploring the leading candidates, only some of which have been
given in Table~\ref{table:results}. Those interested are
encouraged to contact PJL and PJS.

\begin{acknowledgments}
We thank the ICDD for providing a copy of the ICDD-PDF; P. Tandy
(Natural History Museum, London), J. Post (Smithsonian) and  L.
Cabri (CANMET) for lending us mineral samples for pilot studies;
and A.-P. Tsai for sending us the CdYb powder pattern. PJL thanks
B. Fujito and M. Yogo for assistance with coding and statistics,
respectively. This work was supported in part by the US Department
of Energy grant DE-FG02-91ER40671 (PJS) and NSF Materials Research
and Engineering Center grant DMR-94-00362 (NY).
\end{acknowledgments}

\bibliography{peterlu1}

\end{document}